\documentclass[a4paper,12pt]{article}
\pdfoutput=1
\usepackage{graphicx}
\usepackage{mathtools}
\usepackage{amssymb}
\usepackage{amsfonts}
\usepackage[footnotesize]{caption}
\usepackage{color}
\usepackage{braket}
\usepackage{cite}
\usepackage{hyperref}
\hypersetup{pageanchor=false}
\usepackage{url}
\usepackage{multirow}
\usepackage{relsize}
\usepackage{fullpage}
\usepackage{makecell}
\usepackage{fullpage}
\usepackage{blkarray}
\usepackage{placeins}

\usepackage[utf8]{inputenc}

\newcommand{\overbar}[1]{\mkern 1.5mu\overline{\mkern-1.5mu#1\mkern-1.5mu}\mkern 1.5mu}
\newcommand{\pmatr}[1]{\begin{pmatrix} #1 \end{pmatrix}}
\newcommand{\simlt}{~\mbox{\smaller\(\lesssim\)}~}
\newcommand{\simgt}{~\mbox{\smaller\(\gtrsim\)}~}

\newcommand{\exto}[1]{{\smaller$ \times 10^{#1} $}}
\newcommand{\degr}{$^\circ$}
\newcommand{\diag}{\mathrm{diag}}

\begin{document}

\begin{titlepage}
\begin{center}
{\bf\Large Leptogenesis in a \boldmath{$\Delta(27) \times SO(10)$} SUSY GUT} \\[12mm]
Fredrik~Bj\"{o}rkeroth$^{\star}$%
\footnote{E-mail: {\tt f.bjorkeroth@soton.ac.uk}},
Francisco~J.~de~Anda$^{\dagger}$%
\footnote{E-mail: \texttt{franciscojosedea@gmail.com}},
Ivo~de~Medeiros~Varzielas$^{\star}$%
\footnote{E-mail: \texttt{ivo.de@soton.ac.uk}},
Stephen~F.~King$^{\star}$%
\footnote{E-mail: \texttt{king@soton.ac.uk}}
\\[-2mm]

\end{center}
\vspace*{0.50cm}
\centerline{$^{\star}$ \it
School of Physics and Astronomy, University of Southampton,}
\centerline{\it
SO17 1BJ Southampton, United Kingdom }
\vspace*{0.2cm}
\centerline{$^{\dagger}$ \it
Departamento de F{\'i}sica, CUCEI, Universidad de Guadalajara, M{\'e}xico}
\vspace*{1.20cm}

\begin{abstract}
{\noindent
Although $SO(10)$ Supersymmetric (SUSY) Grand Unification Theories (GUTs) are very attractive for neutrino mass and mixing, it is often quite difficult to achieve successful leptogenesis from the lightest right-handed neutrino $N_1$ due to the strong relations between neutrino and up-type quark Yukawa couplings.
We show that in a realistic model these constraints are relaxed, making $N_1$ leptogenesis viable.
To illustrate this, we calculate the baryon asymmetry of the Universe $ Y_B $ from flavoured $ N_1 $ leptogenesis in a recently proposed $ \Delta(27) \times SO(10) $ SUSY GUT.
The flavoured Boltzmann equations are solved numerically, and comparison with the observed $ Y_B $ places constraints on the allowed values of right-handed neutrino masses and neutrino Yukawa couplings.
The flavoured $SO(10)$ SUSY GUT is not only fairly complete and predictive in the lepton sector, but can also explain the BAU through leptogenesis with natural values in the lepton sector albeit with some tuning in the quark sector.
}
\end{abstract}
\end{titlepage}

\section{Introduction}

The Standard Model (SM), while otherwise phenomenologically successful, fails to explain the observed baryon asymmetry of the Universe (BAU), i.e. the presence of more matter than antimatter.
The necessary ingredients for producing the BAU have been listed by Sakharov \cite{Sakharov:1967dj}, and even though they are all present in the SM, including $CP$ violation, the amount of asymmetry calculated is off by orders of magnitude \cite{Kuzmin:1985mm}. Therefore, new sources of $CP$ violation are required to understand the BAU, which is measured either with respect to photon density 
\begin{equation}
	\frac{n_B}{n_\gamma} = (6.10 \pm 0.04) \times 10^{-10},
\end{equation} 
or with respect to the entropy density, 
\begin{equation}
	Y_B = (0.87 \pm  0.01) \times 10^{-10},
\end{equation}
see e.g. \cite{DiBari:2012fz} for reviews and \cite{Ade:2015xua} for a recent determination of the error.

One possible new source that is rather well motivated within the leptonic sector is leptogenesis \cite{Fukugita:1986hr,lepto}. In simple extensions of the SM that explain the light neutrino masses by adding heavy right-handed (RH) Majorana neutrinos \cite{Minkowski:1977sc}, known as the (type I) seesaw mechanism, the decays of the heavy neutrinos can lead to a lepton asymmetry which can be converted into the BAU by nonperturbative SM sphaleron interactions. 
Leptogenesis is a generic feature of models with heavy RH neutrinos and seesaw. Whether or not it can explain the BAU is a \emph{quantitative} problem.
Thermal leptogenesis, wherein the initial RH neutrino abundance is assumed zero (produced instead in the thermal bath), is a particularly minimal incarnation of this mechanism.

Conventional wisdom when discussing leptogenesis in $SO(10)$ \cite{Georgi:1974my} suggests that the lightest RH neutrino $ N_1 $ has a mass that is too low to produce the correct baryon asymmetry. It can be understood as follows: there is a very strong hierarchy in the up-type quark masses, with $ m_u : m_c : m_t \sim 10^{-5} : 10^{-3} : 1 $, while the hierarchy among neutrinos is comparatively mild. Assuming a normal ordering $ m_1 < m_2 < m_3 $, we have $ m_1 : m_2 : m_3 \sim 10^{-2} : 10^{-1} : 1 $. If up-type quark and neutrino Dirac couplings are assumed equal in naive $ SO(10) $, producing the correct hierarchy in the neutrino Majorana masses after seesaw requires a large hierarchy in the RH neutrino masses $ M_i $, like $ 10^6 : 10^{10} : 10^{15} $. Since $M_1$ is too light for traditional leptogenesis, 
typically one proceeds by considering $ N_2 $ leptogenesis \cite{DiBari:2005st}, which has been studied in detail for $ SO(10) $-inspired models \cite{DiBari:2014eya} (for further work on leptogenesis in $ SO(10) $, see \cite{Smirnov:1993af}). 

The situation can be different in $SO(10)$ flavour models \cite{deMedeirosVarzielas:2006fc,SO10}.
In this paper, we will show that $ N_1 $ leptogenesis is possible in a realistic flavoured $SO(10)$ SUSY GUT model,
involving a flavour symmetry to account for the mass hierarchies and neutrino mass and mixing pattern.
As a concrete example, we estimate the BAU arising from leptogenesis in a 
$\Delta(27) \times SO(10)$ SUSY GUT model \cite{Bjorkeroth:2015uou}, which was shown to successfully and accurately fit all quark and lepton mass and mixing parameters, while simultaneously resolving the doublet-triplet splitting problem and demonstrating that proton decay is naturally suppressed well within the current experimental constraints. It further predicts normal neutrino ordering, and a leptonic Dirac phase $ \delta_\mathrm{CP} \approx -\pi/2 $, in good agreement with current experimental bounds. The model involves only ``named'' representations of $ SO(10) $, i.e. the singlet, fundamental, spinor and adjoint representations. 

A compelling feature of the model is that the mass matrices in each sector (including the light neutrinos after the seesaw has been implemented) have the same universal structure, and the phases and mixing angles in the leptonic sector are guided by the flavour symmetry. This leads also to a rather predictive scenario for leptogenesis, which ultimately allows to constrain some of the free parameters of the model (and indirectly, the mass of the RH neutrinos) in order to obtain the correct asymmetry.

The paper is organized as follows. In Section \ref{sec:model} we briefly summarize the model in \cite{Bjorkeroth:2015uou}.
In Section \ref{seesaw} we show how the seesaw mechanism is implemented in this model.
In Section \ref{sec:lepto} we describe how the BAU is obtained from leptogenesis, and plot the results of solving the Boltzmann equations for the flavoured asymmetries $ Y_{\Delta_\alpha} $. Section \ref{sec:conc} concludes. Appendix \ref{sec:quark} contains additonal information about the quark sector of the model in \cite{Bjorkeroth:2015uou}.
Appendix \ref{sec:yukawa} relates the effective seesaw parameters in the model to the underlying model parameters.
Appendix \ref{sec:seesaw} derives a useful result for the seesaw mechanism with rank-one matrices.

\section{The Model \label{sec:model}}

We begin by giving a summary of the model fully described in \cite{Bjorkeroth:2015uou}. The model aims to be fairly complete and is therefore lengthy. In this paper we will describe only those fields and couplings relevant to analysing leptogenesis. The model is a supersymmetric flavour GUT model, where the $ SO(10) $ gauge group is coupled to a $\Delta(27)\times \mathbb{Z}_{9}\times \mathbb{Z}_{12} $ flavour symmetry, with a $ \mathbb{Z}_4^R $ $ R $-symmetry \cite{Lee:2010gv}. The non-Abelian discrete group $\Delta(27)$  is responsible for the flavour structure,%
\footnote{$\Delta(27)$ was first used as a flavour group in \cite{deMedeirosVarzielas:2006fc,Ma:2006ip}, see also \cite{deMedeirosVarzielas:2006bi}.}
while the Abelian discrete group $\mathbb{Z}_{9}$ works as a shaping symmetry and helps to give correct flavour structure in the masses. A further $\mathbb{Z}_{12}$ symmetry is required to fix the flavon alignment potential and plays no role in the sector relevant to leptogenesis.
Furthermore, the model is $CP$ symmetric and renormalisable at high energies; all nonrenormalisable terms appear after integrating out heavy messenger fields. The model is fairly complete and is built using only small, ``named'' representations $SO(10)$: singlet, fundamental, spinor and adjoint.

The full model contains a large number of fields. We list the ones relevant to our discussion in Table \ref{tab:funfields}, all chiral supermultiplets. The superfield $\Psi$ contains the full SM fermion content plus three RH neutrinos. $H_{10}^{u,d}$ contain each of the MSSM Higgs doublets respectively. Two different $SO(10)$ multiplets are needed to obtain a non-trivial Cabibbo-Kobayashi-Maskawa (CKM) matrix. The superfield $H_{\overbar{16}}$ obtains a vacuum expectation value (VEV) that breaks $SO(10)\to SU(5)$ and gives Majorana masses to RH neutrinos. $H_{45}$ breaks $SU(5)\to SM$ and its VEV is aligned in such a way that it achieves doublet-triplet splitting through the Dimopoulos-Wilczek mechanism \cite{DW}. It also provides the necessary Clebsch-Gordan coefficients to obtain correct quark masses. 
The field $\xi$ breaks $\mathbb{Z}_9$ completely and provides an explanation for the hierarchy between fermion masses \`a la Froggatt-Nielsen. Finally, the $\bar{\phi}$ superfields are flavons that break $\Delta(27)$ and provide the structure of the mass matrices in the so-called CSD3 alignment where 
\begin{equation}
\braket{\overbar{\phi}_\mathrm{atm}} = v_\mathrm{atm} \pmatr{0\\1\\1} , \qquad \braket{\overbar{\phi}_\mathrm{sol}} = v_\mathrm{sol} \pmatr{1\\3\\1} , \qquad \braket{\overbar{\phi}_\mathrm{dec}} = v_\mathrm{dec} \pmatr{0\\0\\1}.
	\label{eq:csd3alignments}
\end{equation}
CSD refers to Constrained Sequential Dominance \cite{King:2005bj}. The viability of the CSD($n$) class of flavour models for explaining neutrino data has been studied in \cite{King:2013iva,Bjorkeroth:2015ora} and leptogenesis in these scenarios was studied in \cite{Bjorkeroth:2015tsa}. The $ SO(10) $ model is successfully fitted to available quark and lepton mass and mixing data, and summarised in Tables \ref{tab:fitoutq}-\ref{tab:fitinl} in Appendix \ref{sec:quark}.

\begin{table}
\centering
\footnotesize
\begin{tabular}{| c | c@{\hskip 5pt}c | c c c |}
\hline
\multirow{2}{*}{\rule{0pt}{4ex}Field}	& \multicolumn{5}{c |}{Representation} \\
\cline{2-6}
\rule{0pt}{3ex}			& $\Delta(27)$ & $SO(10)$ & $\mathbb{Z}_{9}$ &$\mathbb{Z}_{12}$ & $\mathbb{Z}_4^R$ \\ [0.75ex]
\hline \hline
\rule{0pt}{3ex}%
$\Psi$ 			& 3 & 16 & 0 & 0 &1\\
\rule{0pt}{3ex}%
$H_{10}^u$ & 1 & 10 & $6$ & 0 & 0\\
$H_{10}^d$ & 1 & 10 & $5$ & 0& 0\\
$H_{45}$ & 1 & 45 & 0 & 0&0\\
$H_{\overbar{16}}$ & 1 & $\overbar{16}$ & $6$ & 0&0\\
\rule{0pt}{3ex}%
$\overbar{\phi}_\mathrm{dec} $ &  $ \overbar{3} $ & 1 & $6$ & 0&0\\ 
$\overbar{\phi}_\mathrm{atm} $ &  $ \overbar{3} $ & 1 & $1$ & 0 &0\\ 
$\overbar{\phi}_\mathrm{sol} $ &  $ \overbar{3} $ & 1 & $5$ & 6&0\\ 
$\xi$ & 1 & 1 & $1$ & 0&0\\[0.5ex]
\hline
\end{tabular}
\caption{Superfields that appear in the Yukawa superpotential, and their associated charges under the symmetries of the model.}
\label{tab:funfields}
\end{table}

Here we will mostly restrict ourselves to study the sector of the model responsible for neutrino and charged lepton masses. These come from the following superpotential terms
\begin{equation}
\medmuskip=3mu
\thinmuskip=0mu
\thickmuskip=0mu
\begin{split}
	\mathcal{W}_Y^0 &= \Psi_i \Psi_j H_{10}^u 
	\left[ \overbar{\phi}^{i}_\mathrm{dec}\overbar{\phi}^{j}_\mathrm{dec}  \frac{\tilde\lambda^{(u)}_{\mathrm{dec}}}{ M_\chi^{2}}
	+ \overbar{\phi}^{i}_\mathrm{atm} \overbar{\phi}^{j}_\mathrm{atm} \xi  \frac{\tilde\lambda^{(u)}_{\mathrm{atm}}}{ M_\chi^{3}}  + \overbar{\phi}^{i}_\mathrm{sol} \overbar{\phi}^{j}_\mathrm{sol} \xi^2\frac{\tilde\lambda^{(u)}_{\mathrm{sol}}}{ M_\chi^{4}} \right] 
	\\
	&\quad + \Psi_i \Psi_j H_{10}^d 
	\left[ \overbar{\phi}^{i}_\mathrm{dec} \overbar{\phi}^{j}_\mathrm{dec}\xi  \frac{\tilde\lambda^{(d)}_{\mathrm{dec}}}{M_\chi^{3}}
	+ \overbar{\phi}^{i}_\mathrm{atm} \overbar{\phi}^{j}_\mathrm{atm} \xi^2 \frac{\tilde\lambda^{(d)}_{\mathrm{atm}}}{ M_\chi^{4}}  + \overbar{\phi}^{i}_\mathrm{sol} \overbar{\phi}^{j}_\mathrm{sol} \xi^3  \frac{\tilde\lambda^{(d)}_{\mathrm{sol}}}{M_\chi^{5}} \right] \\
	&\quad + \Psi_i \Psi_j H_{\overbar{16}}H_{\overbar{16}} 
	\left[
	\overbar{\phi}^{i}_\mathrm{dec} \overbar{\phi}^{j}_\mathrm{dec} \xi^3\frac{\tilde\lambda^{(M)}_{\mathrm{dec}}}{M_\chi^2 M_{\Omega_\mathrm{dec}}^4} 
	+ \overbar{\phi}^{i}_\mathrm{atm} \overbar{\phi}^{j}_\mathrm{atm} \xi^4 \frac{\tilde\lambda^{(M)}_{\mathrm{atm}}}{M_\chi^3 M_{\Omega_\mathrm{atm}}^4}
	+ \overbar{\phi}^{i}_\mathrm{sol} \overbar{\phi}^{j}_\mathrm{sol} \xi^5 \frac{\tilde\lambda^{(M)}_{\mathrm{sol}}}{M_\chi^4 M_{\Omega_\mathrm{sol}}^4}\right],
\end{split}
\label{eq:WY}
\end{equation}
where the $\tilde{\lambda}$'s are real dimensionless couplings 
(defined in Appendix \ref{sec:quark})
and the indices $i=1,2,3$ cover the 3 generations of triplets (lower indices) and anti-triplets (upper indices) under the $\Delta(27)$ symmetry. The above nonrenormalisable superpotential is the result of integrating out heavy messengers of mass $M_{\chi,\Omega}$ as specified in the full model. 
The terms in the first line provide Dirac mass terms for neutrinos and up-type quarks. 
Those in the second line provide masses for charged leptons and down-type quarks. 
The third line contains Majorana masses for RH neutrinos and we assume the hierarchy between distinct messengers $ M_{\Omega_\mathrm{dec}} < M_{\Omega_\mathrm{atm}},M_{\Omega_\mathrm{sol}} $, such that one RH neutrino is so heavy that it may be considered effectively decoupled.

A short description of the quark sector is available in Appendix \ref{sec:quark}.
The full $\mathcal{W}_Y$ actually contains additional terms where every $M_\chi$ mass insertion can be replaced with the VEV $\braket{H_{45}}$, as shown in Appendix \ref{sec:quark}. 
The effect of including the $ \braket{H_{45}} $ is to introduce Clebsch-Gordan factors which, for a general alignment, leads to different quark and lepton masses.
Without $\braket{H_{45}}$, the up-type quark and neutrino Yukawa couplings would be equal, just as they are in naive $SO(10)$ models.%
\footnote{
In the original model in \cite{Bjorkeroth:2015uou}, it was assumed that $ H_{45} $ gets a VEV aligned in such a way that it only couples to quarks and not to leptons. It is now understood that such an alignment does not exist, and $ H_{45} $ must couple to leptons also. This would spoil the model's precise prediction of the phases in the lepton mass matrices unless we assume $ \braket{H_{45}} $ is real. A real $ \braket{H_{45}} $ leaves the conclusions in \cite{Bjorkeroth:2015uou} for the lepton sector completely unchanged, as the additional terms involving $ \braket{H_{45}} $ can be accounted for by a redefinition of available free parameters $ y^e_\mathrm{atm,sol,dec} $ (defined below in Eqs.~\ref{eq:matrixstructure} and \ref{eq:ydefinitions}). However, it fixes all (but one) phases in the quark sector, which were assumed free. We performed a new fit to the quark parameters in this more predictive setup, and found a good fit to data. The results are given in Tables \ref{tab:fitoutq}-\ref{tab:fitinq} in Appendix \ref{sec:quark}.
}

As a consequence, the quark and lepton mass matrices will have the same CSD3 structure but have different dimensionless parameters (although naturally they are expected to be of the same order). This accounts for the observed differences between quark and charged lepton/neutrino masses. 
The quark sector is largely not relevant for leptogenesis calculations, with the exception of the top mass $ m_t $, which appears when $ \Delta L = 1 $ scatterings like $ q t \rightarrow H \rightarrow \ell N $ are taken into account.

\section{The Seesaw Mechanism}
\label{seesaw}
We first present a simple heuristic argument which shows that the seesaw mechanism leads to a light Majorana
neutrino mass matrix with the same universal structure as the input Yukawa and heavy Majorana mass matrices,
then demonstrate this result rigorously.
Below the $SO(10)$ breaking scale,
the flavour structure of the operators emerging from Eq.~\ref{eq:WY} relevant for the neutrino sector may be written schematically as
\begin{equation}
\begin{split}
	&  H_u
	(\overbar{\phi}_\mathrm{atm} L)(\overbar{\phi}_\mathrm{atm} N^c) 
	+ 
	H_u
	(\overbar{\phi}_\mathrm{sol} L)( \overbar{\phi}_\mathrm{sol}N^c) 
	+
	 H_u
	( \overbar{\phi}_\mathrm{dec}L)( \overbar{\phi}_\mathrm{dec}N^c) \\
	& \quad + 
	( \overbar{\phi}_\mathrm{atm}N^c)( \overbar{\phi}_\mathrm{atm}N^c) 
	+
	( \overbar{\phi}_\mathrm{sol}N^c)( \overbar{\phi}_\mathrm{sol}N^c) 
	+
	( \overbar{\phi}_\mathrm{dec}N^c)( \overbar{\phi}_\mathrm{dec}N^c),
\end{split}
\label{eq:lambdanu}
\end{equation}
where we have dropped all the couplings and mass scales, and have not distinguished between flavon fields and their VEVs, all these details will be recovered later.

Noting that the same combinations of RH neutrinos $(\overbar{\phi}_\mathrm{atm} N^c)$, $(\overbar{\phi}_\mathrm{sol} N^c)$, 
$(\overbar{\phi}_\mathrm{dec} N^c)$ appear both in the Dirac and heavy Majorana sectors, integrating out these
combinations of heavy RH neutrinos leads to effective Weinberg operators of the form,
\begin{equation}
	H_uH_u(\overbar{\phi}_\mathrm{atm} L)(\overbar{\phi}_\mathrm{atm} L)
	+H_uH_u (\overbar{\phi}_\mathrm{sol} L)(\overbar{\phi}_\mathrm{sol} L)
	+H_uH_u (\overbar{\phi}_\mathrm{dec} L)(\overbar{\phi}_\mathrm{dec} L),
	\label{eq:lambdanu2}
\end{equation}
which have the same flavour structure as in the original Dirac and heavy Majorana sectors.
According to this heuristic argument one expects that the light effective Majorana neutrino masses will have the same universal structure
as all the other mass matrices, which is an attractive feature of the model.
This simple argument was first presented for the case of tri-bimaximal mixing in \cite{deMedeirosVarzielas:2006fc,deMedeirosVarzielas:2005ax}. However one may worry that the combinations of RH neutrinos that are integrated out, namely
$(\overbar{\phi}_\mathrm{atm} N^c)$, $(\overbar{\phi}_\mathrm{sol} N^c)$, $(\overbar{\phi}_\mathrm{dec} N^c)$ are not mass eigenstates.
One may also worry that the mechanism only works for tri-bimaximal mixing where the flavon alignments are mutually orthogonal.

We now present a more rigorous discussion of the seesaw mechanism in this model, showing that the above result is in fact robust.
From the above superpotential terms one can write the fermionic part of the seesaw Lagrangian as
\begin{equation}
	\mathcal{L} =
		H_u \overline{L}_iY^\nu_{ij} N_{j}
		+ H_d \overline{L}_i Y^e_{ij}e_j 
		+ N_{i}^{T} M^N_{ij} N_{j},
\label{eq:smm}
\end{equation}
where $ L_i $ and $ H_{u,d} $ are the $ SU(2) $ lepton and Higgs doublets, $N_{j}$ are the RH neutrinos, and the Yukawa and Majorana mass
matrices are given in terms of fundamental model parameters in Appendix~\ref{sec:yukawa}.
The Yukawa and mass matrices in Eq.~\ref{eq:Yflav} may be written as%
\footnote{Throughout this paper we use the notation $y_\mathrm{dec}^{e,\nu}, y_\mathrm{sol}^{e,\nu}, y_\mathrm{atm}^{e,\nu}$ and $M_\mathrm{dec}, M_\mathrm{sol}, M_\mathrm{atm}$ which references the flavons involved in the respective Yukawa and Majorana mass terms shown in Eq.~\ref{eq:Yflav}. We note that this differs from the notation used in \cite{Bjorkeroth:2015uou}, where the corresponding parameters for up ($ u $), down ($ d $) and charged lepton ($e$) Yukawa matrices were denoted $y_3^{f}, y_2^{f}, y_1^{f}$, with $ f = u, d, e $.}
\begin{equation}
\begin{split}
	Y^{e,\nu}&=
	y_\mathrm{atm}^{e,\nu} \pmatr{ 0&0&0 \\ 0&1&1 \\ 0&1&1 } 
	+ y_\mathrm{sol}^{e,\nu} e^{i\eta} \pmatr{ 1&3&1 \\ 3&9&3 \\ 1&3&1 }
	+ y_\mathrm{dec}^{e,\nu} e^{i\eta^\prime} \pmatr{ 0&0&0 \\ 0&0&0 \\ 0&0&1 } , \\
	M^N&=
	M_\mathrm{atm} \pmatr{ 0&0&0 \\ 0&1&1 \\ 0&1&1 } 
	+ M_\mathrm{sol} e^{i\eta} \pmatr{ 1&3&1 \\ 3&9&3 \\ 1&3&1 }
	+ M_\mathrm{dec} e^{i\eta^\prime} \pmatr{ 0&0&0 \\ 0&0&0 \\ 0&0&1 } ,
\end{split}
\label{eq:matrixstructure}
\end{equation}
where the real Yukawa parameters and Majorana masses introduced above are given in terms of the fundamental model parameters in Appendix~\ref{sec:yukawa}.
The model fixes $ \eta = 2\pi/3 $, $ \eta^\prime = 0 $, while the effective couplings $ y^{e,\nu}_i $ and $ M_i $ (with $i = \mathrm{dec, atm, sol}$) are real and dimensionless with the natural hierarchies 
\begin{equation}
\begin{array}{rcccl}
	y_\mathrm{dec}^{e,\nu} &\gg& y_\mathrm{atm}^{e,\nu} &\gg& y_\mathrm{sol}^{e,\nu} , \\
	M_\mathrm{dec} &\gg& M_\mathrm{atm} &> & M_\mathrm{sol}.
\end{array}
\label{eq:hierarchies}
\end{equation} 
These relations are a direct consequence of the superpotential in Eq.~\ref{eq:WY} and the symmetry breaking sector that fixes the flavon VEVs,
although, apart from these general expectations, we shall regard these as free parameters. 
Using the standard seesaw formula 
\begin{equation}
m^\nu=v_u^2Y^\nu(M^N)^{-1}(Y^\nu)^T,
\end{equation}
leads to the effective light neutrino matrix with the same structure as Eq.~\ref{eq:matrixstructure},
\begin{equation}
	m^\nu=
	\mu_\mathrm{atm} \pmatr{ 0&0&0 \\ 0&1&1 \\ 0&1&1 } 
	+ \mu_\mathrm{sol} e^{i\eta} \pmatr{ 1&3&1 \\ 3&9&3 \\ 1&3&1 }
	+ \mu_\mathrm{dec} e^{i\eta^\prime} \pmatr{ 0&0&0 \\ 0&0&0 \\ 0&0&1 } ,
\label{mnu}
\end{equation}
with the same values of the phases as before, namely $ \eta = 2\pi/3$, $ \eta^\prime = 0$, and where
\begin{equation}
\mu_\mathrm{atm} \equiv \frac{(y^{\nu}_\mathrm{atm} v_u)^2}{M_\mathrm{atm}},\ \ 
\mu_\mathrm{sol} \equiv \frac{(y^{\nu}_\mathrm{sol} v_u)^2}{M_\mathrm{sol}},\ \ 
\mu_\mathrm{dec} \equiv \frac{(y^{\nu}_\mathrm{dec} v_u)^2}{M_\mathrm{dec}}.
\label{eq:mu}
\end{equation}
This remarkable and non-trivial result, that the seesaw mechanism result for $m^\nu$ in Eq.~\ref{mnu}
preserves the universal matrix structure of $Y^{\nu}$ and $M^N$ in Eq.~\ref{eq:matrixstructure},
is nothing to do with symmetry or the special CSD3 form of alignments. The result only requires that the Dirac and heavy Majorana 
matrices be expressable as linear combinations of the same three rank-one matrices, as shown in Appendix~\ref{sec:seesaw}.

We emphasise that the universal structure of all the Yukawa and Majorana mass matrices (both heavy and light) 
is an attractive feature of the model. Moreover,
due to the small number of free parameters, this is a highly predictive setup. In the lepton sector, only six 
effective free parameters (three $ \mu_i $ and three $ y^e_i $) determine the charged lepton and light neutrino masses, as well as all Pontecorvo-Maki-Nakagawa-Sakata (PMNS) matrix parameters, in excellent agreement with experimental best fits.
This is summarised in Appendix \ref{sec:quark} (see e.g. Tables \ref{tab:fitoutl}, \ref{tab:fitinl} where best fit values of $ \mu_i $ are
presented).
The parameters $ \mu_i $ only fix the combinations of neutrino Dirac and Majorana couplings in Eq.~\ref{eq:mu}
and do not allow the three RH neutrino mass parameters $M_\mathrm{atm}$, $M_\mathrm{sol}$, $M_\mathrm{dec}$ to be disentangled from the Yukawa couplings $y^{\nu}_\mathrm{atm}$, $y^{\nu}_\mathrm{sol}$, $y^{\nu}_\mathrm{dec}$. 

In the next section we show how the requirement that the BAU is produced entirely from thermal $ N_1 $ leptogenesis may constrain the 
Yukawa couplings $y^{\nu}_\mathrm{atm}$, $y^{\nu}_\mathrm{sol}$, which enables the RH neutrino mass parameters $M_\mathrm{atm}$, $M_\mathrm{sol}$ to be constrained, as well as the 
lightest two RH neutrino mass eigenvalues $M_1$ and $M_2$, assuming the third mass $M_3$ to be much heavier.
The relation between the RH mass parameters $M_\mathrm{atm}$, $M_\mathrm{sol}$ and the RH mass eigenvalues
$M_1$, $M_2$ is rather complicated since the RH neutrino mass matrices in Eq.~\ref{eq:matrixstructure} are not diagonal, but according to SD we should have $M_\mathrm{dec}\approx M_3$ which is much heavier than the other RH masses and thus essentially decoupled.

\section{Leptogenesis \label{sec:lepto}}

In this section we shall calculate the BAU using flavoured $N_1$ leptogenesis, for the seesaw matrices derived in the previous section
from our model. Unlike typical $SO(10)$-inspired models, in the flavoured $SO(10)$ model considered here, we will show that it is possible
to obtain $M_1$ large enough to allow for successful $N_1$ leptogenesis. 
As we are considering thermal leptogenesis, we additionally assume a (reheating) temperature $ T > M_1 $.
As the model establishes the hierarchy in RH neutrino masses $M_3 \gg M_2 \gg M_1$, we will use the hierarchical approximation
to leptogenesis generated only by the lightest RH neutrino.
Following the procedure outlined in \cite{Antusch:2006cw}, we may parametrise the final baryon asymmetry $ Y_B $ as 
\begin{equation}
	Y_B = 
	\frac{10}{31}
	\left[ Y_{N_1} + Y_{\tilde{N}_1} \right]_{z \ll 1} 
	\sum_{\alpha} \varepsilon_{1,\alpha} \eta_{\alpha}.
\label{eq:BAU}
\end{equation}
$ \varepsilon_{1,\alpha} $ is the decay asymmetry of $ N_1 $ (s)neutrinos, while $ \eta_{\alpha} $ is an efficiency factor, which contains the dependence on washout from inverse decays and scattering, and is typically different for each flavour $ \alpha $.
In the fully flavoured regime, calculating $ \eta_\alpha $ requires solving the Boltzmann equations in terms of the decay factors $ K_\alpha $ and a numerical $ 3\times3 $ matrix $ A $ that describes flavour coupling effects. 
As can be seen in \cite{Antusch:2006cw}, $ \eta_\alpha $ typically takes values $ 0 < \eta_\alpha \simlt 0.2 $. 

The flavoured decay asymmetry $ \varepsilon_{1,\alpha} $ is given by \cite{Abada:2006fw}
\begin{equation}
	\varepsilon_{1,\alpha} = \frac{\Gamma_{1\alpha} - \overbar{\Gamma}_{1\alpha}}{\Gamma_1 + \overbar{\Gamma}_1},
\end{equation}
where $ \Gamma_{1\alpha} $, $ \overbar{\Gamma}_{1\alpha} $ are the decay rates of $ N_1 $ neutrinos decaying, respectively, into $ \ell_\alpha H_u $ lepton-Higgs or $ \overline{\ell}_\alpha H_u^\ast $ antilepton-Higgs pairs, in a given flavour $ \alpha $. $ \Gamma_1 $ and $ \overbar{\Gamma}_1 $ are the corresponding total decay widths (summed over flavour). An analogous decay asymmetry 
$ \varepsilon_{1,\tilde\alpha} $,
may be defined for neutrinos decaying into $ \tilde{\ell}_\alpha \tilde{H} $ slepton-Higgsino pairs, and similarly we may define
$ \varepsilon_{\tilde 1,\alpha} $, and 
$ \varepsilon_{\tilde 1,\tilde\alpha} $ 
for $ \tilde{N}_1 $ sneutrino decays. In the MSSM, to which the $ SO(10) $ model reduces, all these decay rates are equal, i.e. $ \varepsilon_{1,\alpha} = \varepsilon_{1,\tilde\alpha} = \varepsilon_{\tilde 1,\alpha} = \varepsilon_{\tilde 1,\tilde\alpha} $. 

Assuming $ M_3 $ is large enough that the $ N_3 $ neutrino does not affect leptogenesis, in the hierarchical approximation $ M_1 \ll M_2 $, $  \varepsilon_{1,\alpha} $ can be expressed as
\begin{equation}
	\varepsilon_{1,\alpha} = \frac{1}{8\pi} 
	\frac{\mathrm{Im}\left[(\lambda_\nu^\dagger)_{1\alpha}(\lambda_\nu^\dagger\lambda_\nu)_{12}(\lambda_\nu^T)_{2\alpha}\right]}
		{(\lambda_\nu^\dagger \lambda_\nu)_{11}} 
	g^{\mathrm{MSSM}}\left(\frac{M_2^2}{M_1^2}\right), 
	\quad g^{\mathrm{MSSM}}\left(\frac{M_2^2}{M_1^2}\right) \approx - 3 \frac{M_1}{M_2},
\label{eq:epsilon}
\end{equation}
where $\lambda_\nu$ is the neutrino Yukawa matrix in the flavour basis, where the charged lepton and RH neutrino mass matrices are diagonal by definition. In general, the charged lepton and RH neutrino mass matrices may be diagonalised as follows:
\begin{equation}
\begin{split}
	V_{eL}Y^eV_{eR}^\dagger &= \diag(y_e,y_\mu,y_\tau),\\
	V_{eL}Y^{e\dagger}Y^eV_{eL}^\dagger 
	&= \diag(y_e^2,y_\mu^2,y_\tau^2)
	= V_{eR}Y^{e}Y^{e\dagger}V_{eR}^\dagger ,\\
	U_N M^N U_N^T &= \diag(M_1,M_2,M_3).
\end{split}
\end{equation}
As $ Y^e $ in Eq.~\ref{eq:matrixstructure} is complex symmetric, we have $ V_{eR}^\dagger = V_{eL}^T $.
The neutrino Yukawa matrix in the flavour basis is thus given by 
\begin{equation}
\lambda_\nu = V_{eL} Y^\nu U_N^T.
\end{equation}

Given the highly non-trivial mass and Yukawa matrix structure, to rigorously show that $ N_1 $ leptogenesis can be achieved in this model, we solve the Boltzmann equations for the evolution of the $ N_1 $ neutrino and $ B-L $ asymmetry densities.
We may derive bounds on the neutrino Yukawa couplings by performing a scan over parameter space.
Our analysis is based on results in \cite{Abada:2006ea,Antusch:2006cw}, which give the Boltzmann equations for flavoured (supersymmetric) $ N_1 $ leptogenesis.

We will find that the solutions are chiefly dependent on the decay factors $ K_\alpha $ (themselves dependent on the neutrino Dirac matrix) and a matrix $ A $ that describes flavour coupling effects that modify the lepton asymmetries in individual flavours. 
In the three-flavour case, $ K_\alpha $ and the total decay factor $ K $ are defined by
\begin{equation}
	K_\alpha = \frac{v_u^2 (\lambda_\nu^\dagger)_{1\alpha} (\lambda_\nu)_{\alpha1}}{m_\star M_1}, 
	\quad K = \sum_\alpha K_\alpha \,;
	\quad m_\star \simeq 1.58 \times 10^{-3} \mathrm{~eV}, 
\end{equation}
while 
\begin{equation}
	A = \pmatr{-93/110&6/55&6/55\\3/40&-19/30&1/30\\3/40&1/30&-19/30}.
\end{equation}

The $ N_1 $ neutrino density is given by $ Y_{N_1} $, with the density at thermal equilibrium given by $ Y_{N_1}^\mathrm{eq} $. We define $ \Delta Y_{N_1} = Y_{N_1} - Y_{N_1}^\mathrm{eq} $, as well as corresponding $ \Delta Y_{\tilde{N}_1} $ for the sneutrino density $ Y_{\tilde{N}_1} $. 
The equilibrium density for leptons and sleptons are denoted $ Y_{\ell}^\mathrm{eq} $ and $ Y_{\tilde{\ell}}^\mathrm{eq} $.
We use
\begin{equation}
	Y_{N_1}^\mathrm{eq} = Y_{\tilde{N}_1}^\mathrm{eq} \approx \frac{45}{2\pi^4 g_\ast} z^2 K_2(z), \qquad 
	Y_\ell^\mathrm{eq} = Y_{\tilde{\ell}}^\mathrm{eq} \approx \frac{45}{2\pi^4 g_\ast}.
\end{equation}

The total $ B/3 - L_\alpha $ asymmetries (including both fermion and scalar matter) are given by $ Y_{\Delta_\alpha} $.
The Boltzmann equations may be written as
\begin{align}
	\frac{dY_{N_1}}{dz} &= -2 D f_1 \Delta Y_{N_1} , \label{eq:YN1}\\
	\frac{dY_{\tilde{N}_1}}{dz} &= -2 D f_1 \Delta Y_{\tilde{N}_1} , \\
	\frac{dY_{\Delta_\alpha}}{dz} &= 2 \, \varepsilon_{1,\alpha} D f_1 (\Delta Y_{N_1} + \Delta Y_{\tilde{N}_1})  + W \frac{K_\alpha}{K}  f_2 \sum_\beta A_{\alpha\beta} Y_{\Delta_\beta}. \label{eq:YDelta}
\end{align}
The decay and washout terms $ D $ and $ W $ are defined as
\begin{equation}
	D = K z \frac{K_1(z)}{K_2(z)}, \qquad 
	W = K z \frac{K_1(z)}{K_2(z)} \frac{Y_{N_1}^\mathrm{eq}+Y_{\tilde{N}_1}^\mathrm{eq}}{Y_{\ell}^\mathrm{eq}+Y_{\tilde{\ell}}^\mathrm{eq}}.
\end{equation}
The functions $ f_1(z) $ and $ f_2(z) $ parametrise the contributions from $ \Delta L = 1 $ scatterings. We use the results from \cite{Buchmuller:2004nz}, wherein they consider scatterings involving neutrinos and top quarks but not gauge bosons, nor do they consider thermal effects. 
The functions may be approximated by
\begin{equation}
\begin{split}
	f_1(z) &\approx f_2(z) \approx \frac{z}{a} \left[ \ln\left(1+\frac{a}{z}\right) + \frac{K_S}{K z} \right] \left(1+\frac{15}{8z}\right), \\
	a &= \frac{K}{K_S \ln(M_1/M_h)}, \quad \frac{K_S}{K} = \frac{9}{4\pi^2}\frac{m_t^2}{g_{N_1} v^2},
\end{split}
\end{equation}
where $ m_t $ is the top mass (at the leptogenesis scale), $ M_h \approx 125 $ GeV is the Higgs mass, and $ g_{N_1} = 2 $. In the limit where scattering effects are neglected, $ f_1(z) = f_2(z) = 1 $. The top mass is fitted by the $ SO(10) $ model at the GUT scale, at $ m_t = 92.8 $ GeV. Assuming the running between GUT and leptogenesis scales is relatively minor, we use this benchmark value.

\begin{figure}[ht]
	\centering
	\includegraphics[width=0.8\textwidth]{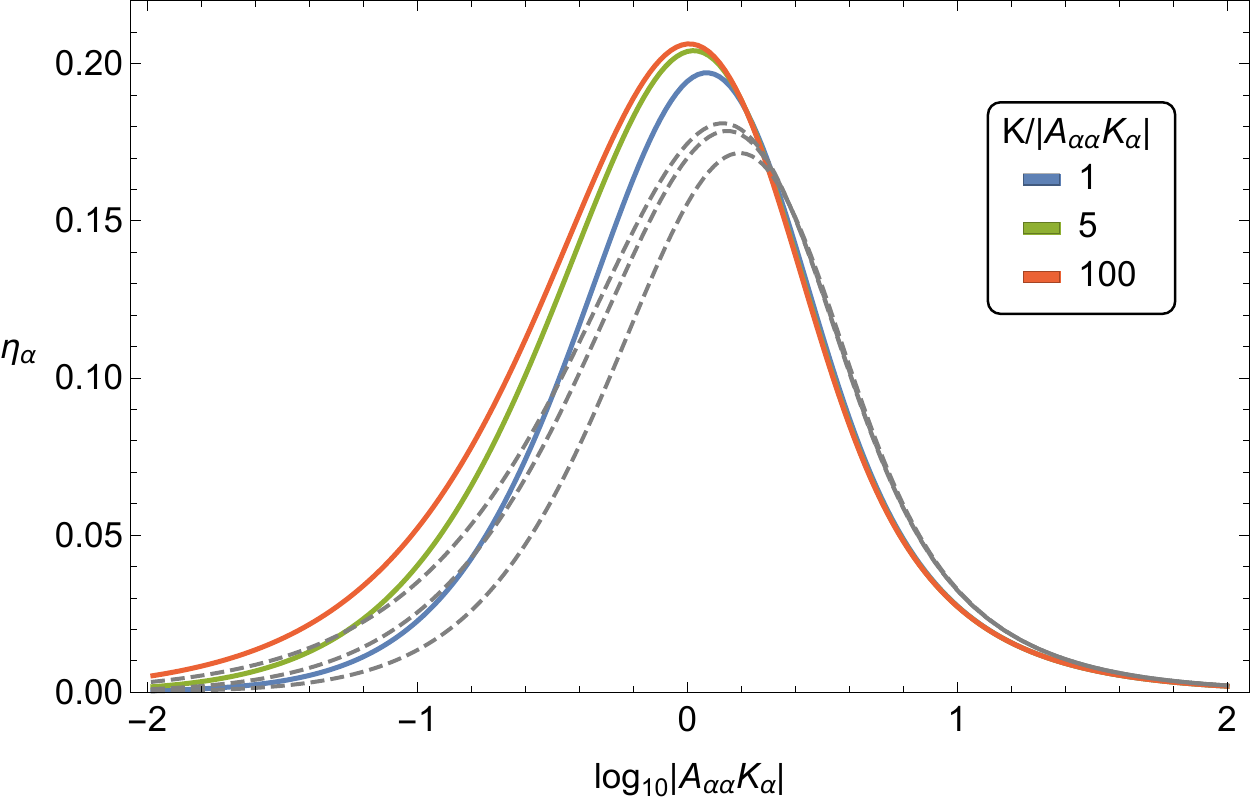}
	\caption{Variation in $ \eta_\alpha $, in agreement with the results in \cite{Antusch:2006cw}. The grey lines show $ \eta_\alpha $ when scatterings are switched off, i.e. $ f_1 = f_2 = 1 $.}
\label{fig:eta}
\end{figure}

We rewrite the parametrisation in Eq.~\ref{eq:BAU} as $ Y_B = Y_0 \sum_\alpha \varepsilon_{1,\alpha} \eta_\alpha $, where $ Y_0 = (10/31)[Y_{N_1} + Y_{\tilde{N}_1}]_{z \ll 1} $ is a normalisation constant that ensures $ 0 \leq \eta_\alpha \leq  1 $. 
We may factor out the decay asymmetry, leading to a set of equations for the efficiency factor $ \eta_\alpha $. 
Furthermore, if we neglect the small off-diagonal elements of the matrix $ A $, the efficiencies in each flavour decouple and may be solved individually, in terms of the decay factors $ K_\alpha $. 
More precisely, for fixed $ K/|A_{\alpha\alpha}K_{\alpha}| $, $ \eta_\alpha (z \rightarrow \infty) $ is a function only of $ A_{\alpha\alpha}K_{\alpha} $. 
Eq.~\ref{eq:YDelta} may be rewritten as
\begin{equation}
	Y_0 \frac{d \eta_\alpha}{dz} = 2 D f_1 (\Delta Y_{N_1} + \Delta Y_{\tilde{N}_1})  + W \frac{A_{\alpha\alpha} K_\alpha}{K}  f_2  Y_0 \eta_\alpha.
\end{equation}
In Fig.~\ref{fig:eta} we show the variation in $ \eta_\alpha $, in agreement with the results in \cite{Antusch:2006cw}. The grey lines show $ \eta_\alpha $ when scatterings are switched off. 

In the solutions presented below, we will solve Eqs.~\ref{eq:YN1}-\ref{eq:YDelta} in terms of the full $ A $-matrix.
The only parameters in our model which are not fixed by the fit to lepton data are either the set of three neutrino Dirac couplings $ y^\nu_i $ or the three RH neutrino Majorana couplings $ M_i $ ($i = \rm{dec}, \rm{atm}, \rm{sol}$). Once either set has been chosen, the other is fixed by the relation $ \mu_i = (v_u y^\nu_i)^2/M_i $. We choose as inputs the Dirac couplings.

Due to the structure of $ SO(10) $, we anticipate these to be roughly equal to the up-type quark Yukawa couplings. As we will find, there exists some tension between the up-quark and neutrino sectors. 
We begin by noting that the third neutrino does not significantly affect the results. 
Thus it is most interesting to examine the $ y^\nu_\mathrm{atm} - y^\nu_\mathrm{sol} $ space, while setting $ y^\nu_\mathrm{dec} = 0.5 $. As a consequence, the third neutrino $ N_3 $ has a mass $ M_3 \approx M_\mathrm{dec} \sim M_\mathrm{GUT} $.

\begin{figure}[ht]
	\centering
	\includegraphics[width=0.9\textwidth]{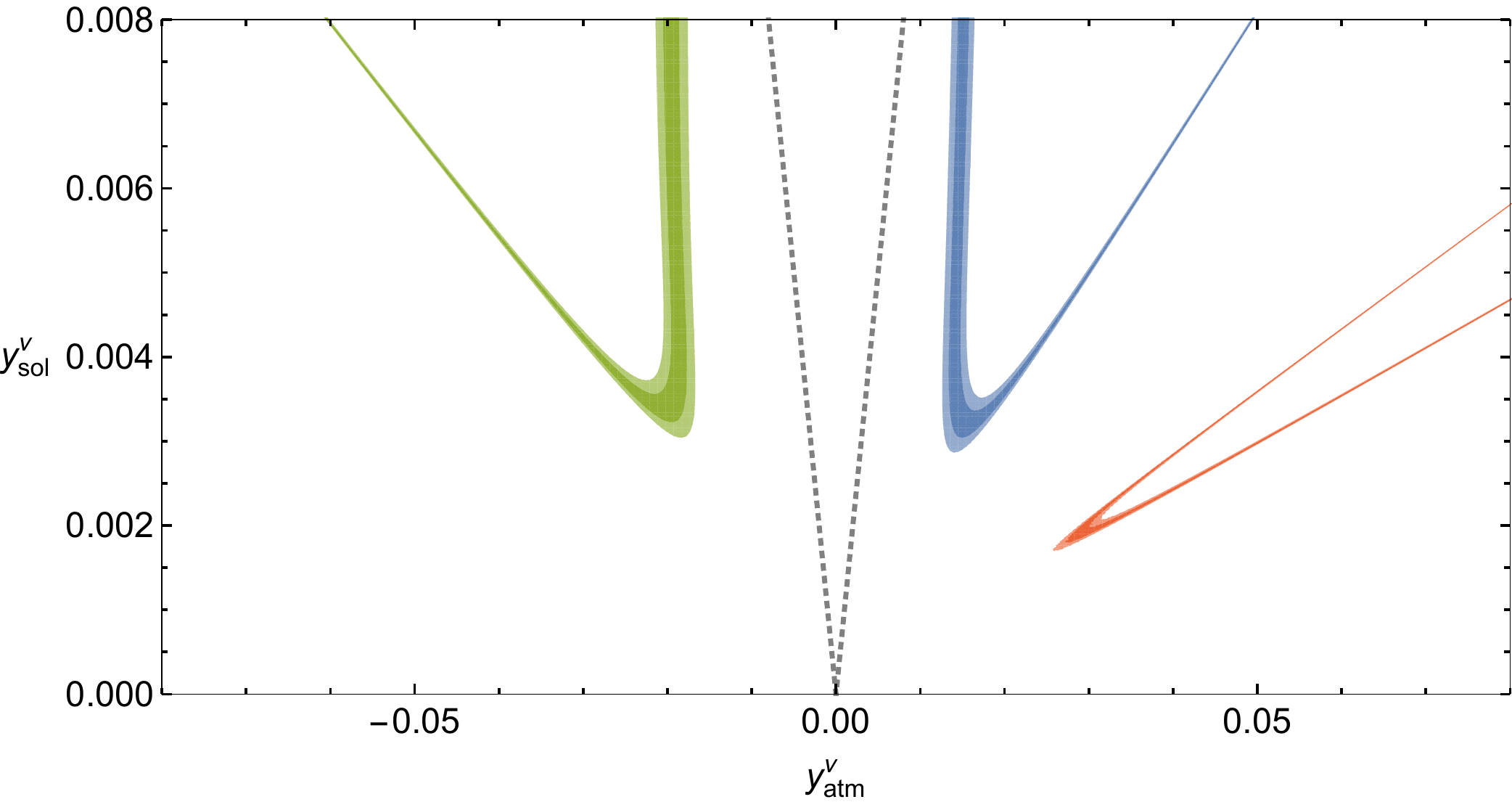}
	\caption{Regions where $ Y_B $ is within 20\% (light bands) and 10\% (darker bands) of the observed BAU, in terms of the neutrino Dirac parameters $ y^\nu_\mathrm{atm} $ and $ y^\nu_\mathrm{sol} $
	where we have assumed $ y^\nu_\mathrm{dec} = 0.5 $. Colours denote separated regions in parameter space. Dotted lines correspond to $ y^\nu_\mathrm{atm} = \pm y^\nu_\mathrm{sol} $.}
\label{fig:ynu1ynu2}
\end{figure}

\begin{figure}[ht]
\centering
	\includegraphics[width=0.7\textwidth]{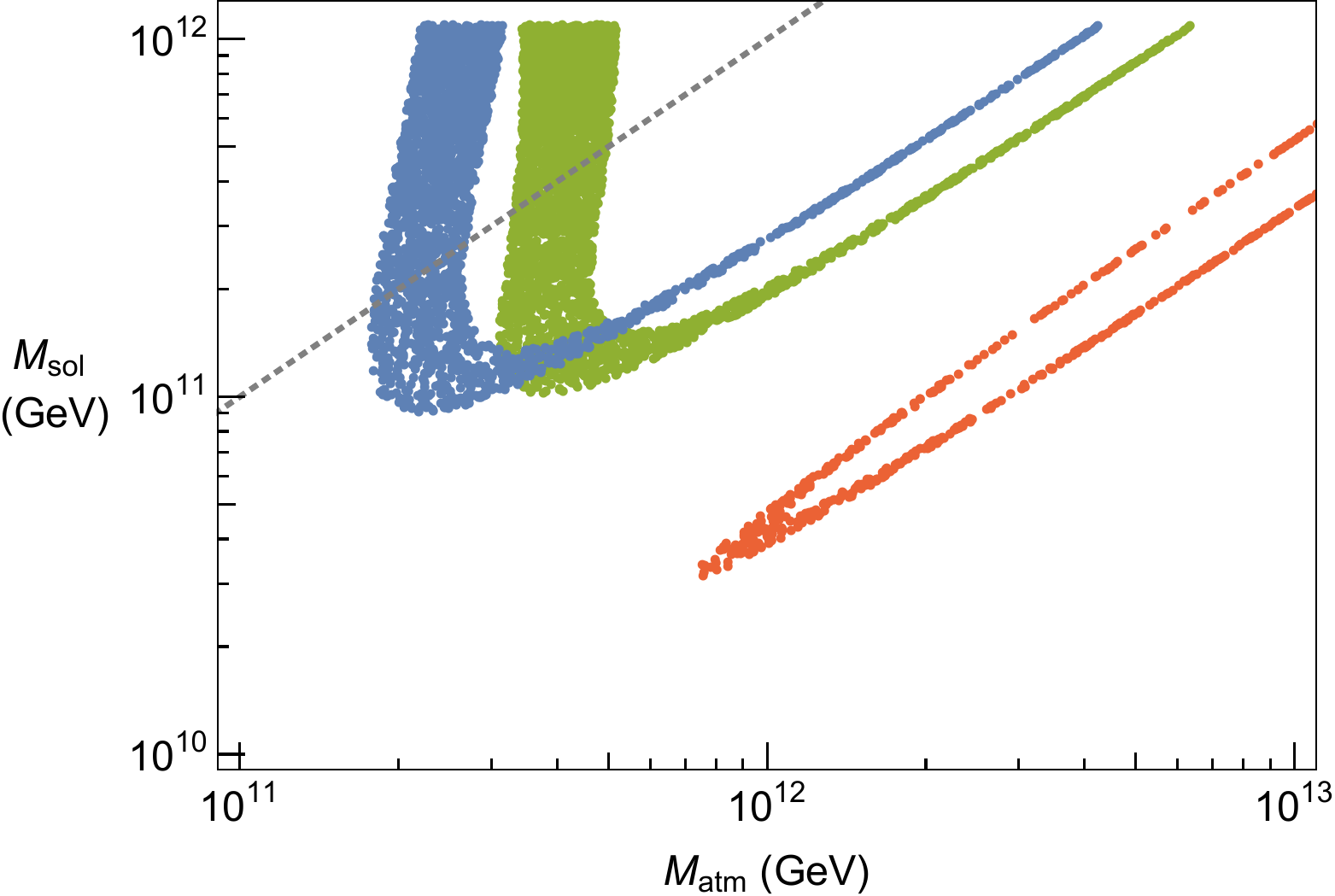}
	\caption{Allowed values of RH neutrino input masses $ M_\mathrm{sol,atm} $, giving $ Y_B $ within 20\% of the observed value, equivalent to the viable choices of Yukawa parameters $ y^\nu_i $ shown in Fig. \ref{fig:ynu1ynu2}. The dotted line corresponds to $ M_\mathrm{atm} = M_\mathrm{sol} $.}
\label{fig:MatmMsol}
\end{figure}

\begin{figure}[ht]
\centering
	\includegraphics[width=0.7\textwidth]{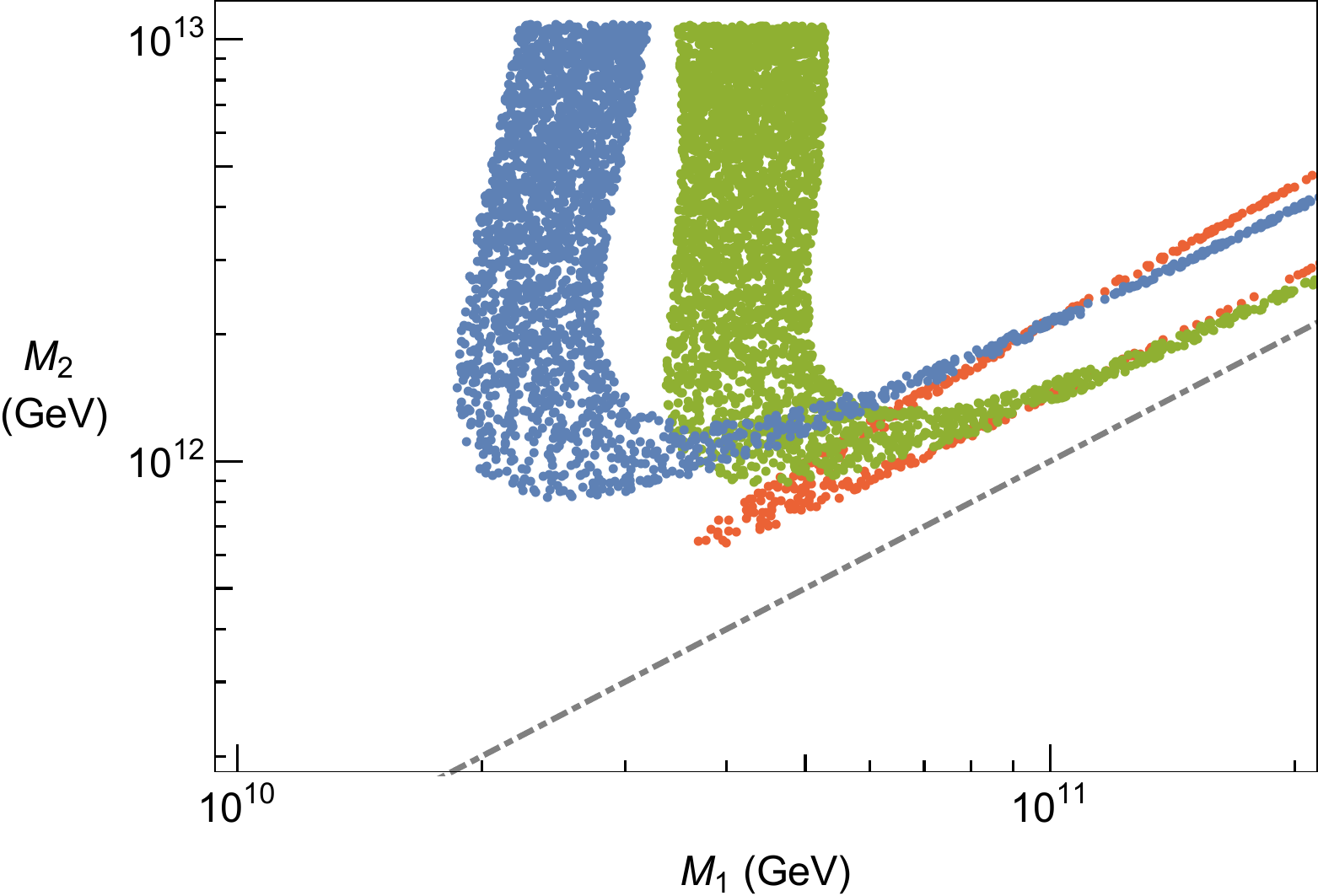}
	\caption{Allowed values of RH neutrino eigenvalues $ M_{1,2}$ (right), giving $ Y_B $ within 20\% of the observed value, equivalent to the viable choices of Yukawa parameters $ y^\nu_i $ shown in Fig. \ref{fig:ynu1ynu2}. The dot-dashed line corresponds to $ M_1 = 0.1 M_2 $.}
\label{fig:M1M2}
\end{figure}

Fig.~\ref{fig:ynu1ynu2} shows the values of the
neutrino Dirac parameters $ y^\nu_\mathrm{atm} $ and $ y^\nu_\mathrm{sol} $
which produce the correct $ Y_B $, to within 10\% and 20\% (darker and lighter shades, respectively) as well as satisfying the phenomenological requirements for correct neutrino masses and lepton mixing.
Each distinct region of parameter space in Fig. \ref{fig:ynu1ynu2} is marked in a different colour, which correlate also with the colours in Figs.~\ref{fig:MatmMsol}-\ref{fig:M1M2}. 
Although the dotted line (indicating $ y_\mathrm{atm} = \pm y_\mathrm{sol} $) in Fig.~\ref{fig:ynu1ynu2} shows that the successful leptogenesis points always satisfy $ y^\nu_\mathrm{atm} > y^\nu_\mathrm{sol} $, the hierarchy is not that strong, bearing in mind that the Dirac Yukawa matrix associated with $ y^\nu_\mathrm{atm} $ in Eq.~\ref{eq:matrixstructure} has numerically smaller entries than that associated with $ y^\nu_\mathrm{sol} $. Consequently both Dirac matrices will contribute significantly to the second column of the total Dirac mass matrix over the successul leptogenesis regions, making any analytic approximation highly non-trivial.

Figs.~\ref{fig:MatmMsol}-\ref{fig:M1M2} show the corresponding RH neutrino mass parameters giving the correct $ Y_B $ to within 20\%, satisfying also the phenomenological requirements for correct neutrino masses and lepton mixing.
Fig.~\ref{fig:MatmMsol} shows input mass parameters $ M_\mathrm{sol,atm} $ while Fig.~\ref{fig:M1M2} shows mass eigenvalues $ M_{1,2} $.
The assumed strong hierarchy $ M_1 \ll M_2 $ is always realised for successful leptogenesis, as shown in Fig.~\ref{fig:M1M2} where all points satisfy $ M_1 < 0.1 M_2 $ (the dot-dashed line shows $ M_1 = 0.1 M_2 $). 

There is however no such strong hierarchy between the mass parameters $ M_\mathrm{sol,atm} $ in Fig.~\ref{fig:MatmMsol}.
Although successful leptogenesis points satisfy $ M_\mathrm{sol} < M_\mathrm{atm} $ over much of parameter space (the dotted line in Fig.~\ref{fig:MatmMsol} marks where $ M_\mathrm{atm} = M_\mathrm{sol} $), it should be noted that the trace of the matrix associated with $ M_\mathrm{sol}$ in Eq.~\ref{eq:matrixstructure} is about five times larger than that associated with $M_\mathrm{atm} $.
We conclude that both these mass matrices will be important in determining the eigenvalues $M_1$ and $M_2$ over the successful leptogenesis regions, and simple approximations are generally not reliable.

We find a lower bound on the parameters giving successful leptogenesis, with $ y^\nu_\mathrm{atm} \simgt 0.01 $ and $ y^\nu_\mathrm{sol} \simgt 0.002 $.
The narrow red region arises from very particular choices of $ y^\nu_i $ that also give the weakest hierarchy of RH neutrino masses $ M_1/M_2 \sim 0.1 $.
Note that the mass $ M_3 $ does not appear in the approximated decay asymmetries and its effect on leptogenesis is always negligible. $ M_3 $ is therefore only constrained by the relation imposed by sequential dominance, i.e. $ M_3 \gg M_{1,2} $.

The effective neutrino couplings $y^\nu_i$ that yield viable leptogenesis (as shown by Fig. \ref{fig:ynu1ynu2}) are within the range anticipated by the model, according to the magnitudes of the VEVs that enter into their definition (see Eq.~\ref{eq:ydefinitions} in Appendix \ref{sec:yukawa}).
We note however that the $y^\nu_i$ are different when compared to the effective up-type quark couplings $y^u_i$ required to obtain correct GUT scale masses $m_u$, $m_c$ and $m_t$.
This would rule out a naive $SO(10)$ model, and while it can be accommodated in this model, there is a price to pay as the $y^\nu_i$ necessarily differ from the $y^u_i$. In particular, $ y^\nu_\mathrm{atm} $ is larger than its corresponding quark parameter by an $ \mathcal{O}(100) $ factor.
This issue is discussed in full detail in Appendix \ref{sec:quark}.


It is interesting to compare this model to another model which incorporates the CSD3 vacuum alignments, based on $ A_4 \times SU(5) $ with two RH neutrinos \cite{Bjorkeroth:2015ora} (see \cite{Bjorkeroth:2015tsa} for a discussion on leptogenesis in this model).
In that model the neutrino Yukawa and RH Majorana mass matrices may be written as
\begin{equation}
	Y^\nu = \pmatr{0 & b \,e^{i \eta/2} 
	\\ a & 3 b \,e^{i \eta/2}
	\\ a & b \,e^{i \eta/2}}, 
	\qquad
	M_R = \pmatr{M_\mathrm{atm}&0\\0&M_\mathrm{sol}},
\label{eq:su5yukawas}
\end{equation}
where $ a $ and $ b $ are real numbers and $ M_\mathrm{atm} = M_1 \ll  M_2 = M_\mathrm{sol} $. 
It was found in this scenario that $ Y_B \propto + \sin \eta $, which gives the correct sign of the asymmetry since the phase is fixed (in both models) to be $ \eta = 2\pi/3 $ by low energy neutrino phenomenology. 

In the present $ SO(10) $ model \cite{Bjorkeroth:2015uou}, let us first consider the regions of parameter space where 
$ y^\nu_\mathrm{atm} \gg y^\nu_\mathrm{sol} $ and $ M_\mathrm{atm} \gg M_\mathrm{sol} $, and the third neutrino is entirely decoupled (i.e. $ M_\mathrm{dec} \rightarrow \infty $). 
In these regions of parameter space, which however do not correspond to the successful leptogenesis regions we have seen, the matrices in Eq.~\ref{eq:matrixstructure} approximate to
\begin{equation}
	Y^\nu \approx \pmatr{y^\nu_\mathrm{sol} \,e^{i \eta} & 3 y^\nu_\mathrm{sol}
	\\ 3 y^\nu_\mathrm{sol} \,e^{i \eta} & y^\nu_\mathrm{atm} 
	\\ y^\nu_\mathrm{sol} \,e^{i \eta} & y^\nu_\mathrm{atm}
	}, 
	\qquad
	M_R \approx \pmatr{M_\mathrm{sol} e^{i \eta} & 0 \\ 0 & M_\mathrm{atm}}.
\label{eq:so10yukawas}
\end{equation}
By the arguments presented in \cite{Antusch:2006cw}, this implies $ Y_B \propto - \sin \eta $,%
\footnote{This can be understood intuitively by noting that Eqs.~\ref{eq:su5yukawas} and \ref{eq:so10yukawas} differ by a column swap in $ Y^\nu $. Under this swap, the relative phase between columns flips sign.}
giving the wrong sign of the asymmetry and an antimatter universe. This is confirmed by the exact numerical solutions which show that these regions of parameter space are not allowed, precisely because they would lead to the wrong sign of the BAU.
The correct sign can be achieved, however, in the regions of parameter space where the above assumptions of a strong hierarchy between `atm' and `sol' are relaxed. These correspond to the successful regions shown in Figs.~\ref{fig:ynu1ynu2} and \ref{fig:MatmMsol}.

We finally note that enforcing the hierarchy $ M_\mathrm{atm} \ll M_\mathrm{sol} $ in the present model, as predicted by the $ SU(5) $ model, does not recover the matrix structure of that model (as seen in Eq.~\ref{eq:su5yukawas}). In this limit, which also requires $ y^\nu_\mathrm{atm} \ll y^\nu_\mathrm{sol} \ll y^\nu_\mathrm{dec} $, the  $ SO(10) $ matrices proportional to $ y^\nu_\mathrm{atm} $ and $ M_\mathrm{atm} $ are negligible, and the total Yukawa and mass matrices approximate to
\begin{equation}
	Y^\nu \approx y^\nu_\mathrm{sol} \,e^{i \eta} 
	\pmatr{ 1&3&1 \\ 3&9&3 \\ 1&3&
		\dfrac{y^\nu_\mathrm{dec}}{y^\nu_\mathrm{sol} e^{i \eta}}
	}
	, \qquad
	M_R \approx M^\nu_\mathrm{sol} \,e^{i \eta} 
	\pmatr{ 1&3&1 \\ 3&9&3 \\ 1&3&
		\dfrac{M^\nu_\mathrm{dec}}{M^\nu_\mathrm{sol} e^{i \eta}}
	},
\label{eq:so10yukawasLSDlimit}
\end{equation}
which are markedly different from the form of Eq.~\ref{eq:su5yukawas}.

\section{Conclusions \label{sec:conc}}

Although $SO(10)$ Grand Unification is a very attractive setting for neutrino mass and mixing, it is often quite difficult to achieve successful leptogenesis from the lightest RH neutrino $N_1$ due to the strong relations between neutrino and up-type quark Yukawa couplings.
In this paper we have shown that a realistic model relaxes these constraints, making $N_1$ leptogenesis viable.

To illustrate this we have calculated the baryon asymmetry of the Universe $ Y_B $ from flavoured $ N_1 $ leptogenesis in a recently proposed $ \Delta(27) \times SO(10) $ SUSY GUT of Flavour.
In this model the lepton mass matrices have a universal structure, and each mass matrix (for charged leptons, Dirac neutrinos and RH neutrinos) is given in terms of three real free parameters, which multiply submatrices that are fixed by the VEVs of $\Delta(27)$ triplet flavons. We have shown that also the mass matrix of the light left-handed neutrinos after  seesaw has this same structure. 
Hence all the low-energy flavour observables are fitted with only 6 real parameters: neutrino mass-squared differences, charged lepton masses and the entire PMNS matrix, a total of 12 parameters (including Majorana phases). This leads to a predictive and highly successful model.

Within the $ \Delta(27) \times SO(10) $ SUSY GUT of Flavour, we have shown that the BAU can be successfully generated via thermal leptogenesis arising from the decay of the lightest RH neutrino. 
The flavoured Boltzmann equations have been solved numerically, and comparison with the observed $ Y_B $ constrains the allowed values of RH neutrino masses and neutrino Yukawa couplings.
The correct BAU can be obtained in a region of parameter space corresponding to RH neutrino masses $M_1 \sim 10^{10-11}$ GeV and $M_2 \sim 10^{11-13}$ GeV, with a strong hierarchy $M_1\ll M_2$ and a temperature $ T > M_1 $. However we have seen that there is no strong hierarchy involving $ M_\mathrm{atm}$, $M_\mathrm{sol} $, i.e. RH neutrino mass eigenstates arise from strongly mixed combinations of $ N^c_\mathrm{atm}$, $N^c_\mathrm{sol} $.

The region of parameter space that has viable BAU has a modest hierarchy between the two relevant neutrino Yukawa couplings, as anticipated by the model. This is in tension with the viable up-type quark Yukawa couplings, which have a stronger hierarchy. The required difference between up and neutrino couplings rules out naive $SO(10)$ models, but can be reconciled in this GUT flavour model, due to the presence of an $ SO(10) $ adjoint field $ H_{45} $.

We conclude that, contrary to expectations based on naive $SO(10)$-inspired assumptions, $N_1$ leptogenesis is viable in flavoured $SO(10)$ SUSY GUTs.
In particular we have shown that the flavoured GUT model detailed in \cite{Bjorkeroth:2015uou} is not only fairly complete and predictive in the lepton sector, but can also explain the BAU through leptogenesis with natural values in the lepton sector albeit with some tuning in the quark sector.

\subsection*{Acknowledgements}
F.B. thanks Stefan Antusch for valuable discussions.
This project has received funding from the European Union's Seventh Framework Programme for research, technological development and demonstration under grant agreement no PIEF-GA-2012-327195 SIFT.
S.\,F.\,K. acknowledges the STFC Consolidated Grant ST/L000296/1 and the European Union's Horizon 2020 Research and Innovation programme under Marie Sk\l{}odowska-Curie grant agreements 
Elusives ITN No.\ 674896 and InvisiblesPlus RISE No.\ 690575.

\FloatBarrier

\appendix

\section{Quark and lepton masses in the model \label{sec:quark}}

In naive $SO(10)$ GUT models, the Yukawa couplings of up-type quarks and neutrinos are the same. As a simple example, in the basis where they are diagonal,
\begin{equation}
	Y^{u} = Y^{\nu} = \pmatr{ y_u & 0 & 0 \\ 0 & y_c & 0 \\ 0 & 0 & y_t }.
\end{equation} 
From this correspondence and the observed values for the light neutrino masses one derives, after seesaw, an expected range for each of the heavy RH neutrino masses $M_{1,2,3}$, using the relation in Eq.~\ref{eq:mu}. This reveals a hierarchy in $ M_i $ like $ 10^6 : 10^{10} : 10^{15} $.

This equality between up-type quarks and neutrinos isn't necessarily in place in all $SO(10)$ models. Indeed, in the model from \cite{Bjorkeroth:2015uou} which we consider here, there are contributions to Yukawa matrices that are common to both up-type quarks and neutrinos, but there are also terms that contribute only to $ Y^u $.

The terms listed in Eq.~\ref{eq:WY} show the effective superpotential after integrating out heavy messengers $ \chi $ at a mass scale $ M_\chi $. In the full superpotential, there are additional terms wherein the messengers $ \chi $ couple to a superfield $H_{45}$ (an $SO(10)$ adjoint) which we assume to acquire a real VEV $ \braket{H_{45}} $. As such, each mass insertion $ M_\chi $ may be replaced by $ \braket{H_{45}} $. The full superpotential is
\begin{equation}
\medmuskip=3mu
\thinmuskip=0mu
\thickmuskip=0mu
\begin{split}
	\mathcal{W}_Y &~=~\Psi_i \Psi_j H_{10}^u 
	\left[ \overbar{\phi}^{i}_\mathrm{dec}\overbar{\phi}^{j}_\mathrm{dec} \sum_{n=0}^2 \frac{\lambda^{(u)}_{\mathrm{dec},n}}{\braket{H_{45}}^n M_\chi^{2-n}}
	+ \overbar{\phi}^{i}_\mathrm{atm} \overbar{\phi}^{j}_\mathrm{atm} \xi \sum_{n=0}^3 \frac{\lambda^{(u)}_{\mathrm{atm},n}}{\braket{H_{45}}^{n} M_\chi^{3-{n}}} \right.
	\\
	& \left. \hspace{14ex} + \overbar{\phi}^{i}_\mathrm{sol} \overbar{\phi}^{j}_\mathrm{sol} \xi^2 \sum_{n=0}^4 \frac{\lambda^{(u)}_{\mathrm{sol},n}}{\braket{H_{45}}^n M_\chi^{4-n}} +\overbar{\phi}^{i}_\mathrm{sol}\overbar{\phi}^{j}_\mathrm{dec}\xi 
	\bigg(\frac{\lambda^{(u)}_{{\rm sd},1}}{\braket{H_{45}'}^2 M_\chi}+\frac{\lambda^{(u)}_{{\rm sd},2}}{\braket{H_{45}'}^2 \braket{H_{45}}}\bigg) \right] 
	\\
	&\quad+ \Psi_i \Psi_j H_{10}^d 
	\left[ \overbar{\phi}^{i}_\mathrm{dec} \overbar{\phi}^{j}_\mathrm{dec}\xi \sum_{n=0}^3 \frac{\lambda^{(d)}_{\mathrm{dec},n}}{\braket{H_{45}}^n M_\chi^{3-n}}
	+ \overbar{\phi}^{i}_\mathrm{atm} \overbar{\phi}^{j}_\mathrm{atm} \xi^2 \sum_{n=0}^4 \frac{\lambda^{(d)}_{\mathrm{atm},n}}{\braket{H_{45}}^{n} M_\chi^{4-{n}}} \right.\\
	&  \left. \hspace{14ex} + \overbar{\phi}^{i}_\mathrm{sol} \overbar{\phi}^{j}_\mathrm{sol} \xi^3 \sum_{n=0}^5 \frac{\lambda^{(d)}_{\mathrm{sol},n}}{\braket{H_{45}}^n M_\chi^{5-n}} \right].
\end{split}
\label{eq:sYW}
\end{equation}

The couplings in Eq. \ref{eq:WY} are then defined in terms of the dimensionless couplings above as
\begin{equation}
\begin{split}
	\frac{\tilde\lambda^{(u,d,M)}_{\rm dec}}{M^2_\chi} &= \sum_{n=0}^2 \frac{\lambda^{(u,d,M)}_{\rm dec,n}}{\braket{H_{45}}^n M_\chi^{2-n}}, \\
	\frac{\tilde\lambda^{(u,d,M)}_{\rm atm}}{M^2_\chi} &= \sum_{n=0}^3 \frac{\lambda^{(u,d,M)}_{\rm atm,n}}{\braket{H_{45}}^n M_\chi^{3-n}}, \\
	\frac{\tilde\lambda^{(u,d,M)}_{\rm sol}}{M^2_\chi} &= \sum_{n=0}^4 \frac{\lambda^{(u,d,M)}_{\rm sol,n}}{\braket{H_{45}}^n M_\chi^{4-n}}.
\end{split}
\end{equation}

The alignment of $ \braket{H_{45}} $ dictates the Clebsch-Gordan coefficients associated with quarks and leptons, which are generally different. For example, if the VEV $\braket{H_{45}}$ is aligned in the $B-L$ direction then $\braket{H_{45}}=v_{45}/3$ for quarks and $\braket{H_{45}}=-v_{45}$ for leptons.
For a general alignment, the practical consequence is that the free parameters in the mass matrices will generally be different for quarks and leptons. However, we assume that $ \braket{H_{45}} $ is real, such that phases only arise from the phases of flavon VEVs, $ \braket{\phi} $ and $ \braket{\xi} $.

In the up sector, there is also an additional set of terms allowed by the symmetries and field content. This is discussed in more detail in \cite{Bjorkeroth:2015uou}. The quark mass matrices can thus be written as
\begin{equation}
\medmuskip=0mu
\thinmuskip=0mu
\thickmuskip=2mu
\begin{split}
	Y^{u} &= 
	y_\mathrm{atm}^{u} \pmatr{ 0&0&0 \\ 0&1&1 \\ 0&1&1 } 
	+ y_\mathrm{sol}^{u} e^{i\eta}\pmatr{ 1&3&1 \\ 3&9&3 \\ 1&3&1 } 
	+ y_\mathrm{dec}^{u} \pmatr{ 0&0&0 \\ 0&0&0 \\ 0&0&1 } 
	+ y_{\rm{sd}}^{u} e^{i\eta^u} \pmatr{ 0&0&1 \\ 0&0&3 \\ 1&3&2 }, \\
	Y^{d} &= 
	y_\mathrm{atm}^{d} \pmatr{ 0&0&0 \\ 0&1&1 \\ 0&1&1 } 
	+ y_\mathrm{sol}^{d} e^{i\eta} \pmatr{ 1&3&1 \\ 3&9&3 \\ 1&3&1 } 
	+ y_\mathrm{dec}^{d} \pmatr{ 0&0&0 \\ 0&0&0 \\ 0&0&1 } .
\end{split}
\end{equation}

The parameters $ y^u_i $ and $ y^\nu_i $ (for each $i = \mathrm{dec}, \mathrm{atm}, \mathrm{sol}$) in the up-type quark and neutrino Yukawa matrices, respectively, come from the same sum of terms involving increasing powers $ \braket{H_{45}} $ in the denominator, but are generally different due to the alignment of $ \braket{H_{45}} $. 
This difference may be parametrised by $\delta_i$, such that
\begin{equation}
y_i^u = y_i^\nu + \delta_i
\label{ydelta}
\end{equation}
There is an additional parameter $y_\mathrm{sd}^{u}$ in the up-type quark matrix. If we were to set these four additional parameters to zero, there would be no difference between the up-type quark and neutrino Yukawa couplings, and the model would follow the expectation of naive $SO(10)$ models: there would be just three independent parameters, $y_\mathrm{dec}^{u}=y_\mathrm{dec}^{\nu}$, $y_\mathrm{atm}^{u}=y_\mathrm{atm}^{\nu}$ and $y_\mathrm{sol}^{u}=y_\mathrm{sol}^{\nu}$ which can be eliminated in terms of the GUT scale values for up, charm and top Yukawa couplings $y_u$, $y_c$, $y_t$.%
\footnote{This is always possible to do numerically, even if analytical relations may become non-trivial from diagonalizing the Yukawa matrices.}

With the additional parameters $\delta_i$ non-zero, $y_i^u$ are related to $y_i^\nu$ as shown in Eq. \ref{ydelta}. The numerical fit to the data indicates (cf. Table~\ref{tab:fitinq}) that $y^u_\mathrm{atm} \sim 10^{-5}$. This is the root of a fine-tuning in the model that arises when we compare it to $y^\nu_\mathrm{atm} $, which, in order to have viable leptogenesis, requires $y^\nu_\mathrm{atm} \sim 10^{-3} - 10^{-2}$, according to Fig.~\ref{fig:ynu1ynu2}. 
This mismatch between up-type quark and neutrino couplings is a typical problem for leptogenesis in $SO(10)$ GUT models, and would invalidate leptogenesis in naive $SO(10)$ models where the couplings need to be equal. 

In the model in question it can be accommodated through a cancellation between $y^\nu_\mathrm{atm}$ and $\delta_\mathrm{atm}$ both of order $10^{-3}-10^{-2}$, leaving $y^u_\mathrm{atm} \sim 10^{-5}$. It should be noted that in the model in question \cite{Bjorkeroth:2015uou}, $y^\nu_\mathrm{atm} \sim 10^{-3} - 10^{-2} $ is indeed the expected order of magnitude for the Dirac neutrino coupling (due to the powers of the superfield $\xi$). It is $y^u_\mathrm{atm} \sim 10^{-5}$ that is required by the fit that turns out anomalously small, which is in turn linked to the mass of the (first generation) up quark, $ m_u $ (see Table \ref{tab:fitoutq}).

The best fit parameters for quarks and leptons are given, respectively, in Tables \ref{tab:fitoutq} and \ref{tab:fitoutl}. The corresponding input parameters are given in Tables \ref{tab:fitinq} and \ref{tab:fitinl}.

\begin{table}[!ht]
\centering
\footnotesize
\renewcommand{\arraystretch}{1.3}
\begin{minipage}[b]{0.50\textwidth}
\centering
\captionsetup{width=\textwidth}
\begin{tabular}{| @{\hskip 8pt}c@{\hskip 6pt}l | c | r@{\hskip 3pt}c@{\hskip 3pt}l |}
\hline
\multicolumn{2}{|c|}{Observables} & \rule{0pt}{4.5ex}Model & \multicolumn{3}{c|}{\makecell{Data fit $ 1\sigma $ range \\ (from \cite{Antusch:2013jca})}} \\[2ex]
\hline\hline
\rule{0pt}{3ex}%
	$ \theta_{12}^q $ & /\degr & 13.020 & 12.985 &$\rightarrow$& 13.067 \\
	$ \theta_{13}^q $ & /\degr & 0.2023 & 0.1866 &$\rightarrow$& 0.2005 \\
	$ \theta_{23}^q $ & /\degr & 2.238 & 2.202 &$\rightarrow$& 2.273 \\
	$ \delta^q $ & /\degr & 69.89 & 66.12 &$\rightarrow$& 72.31 \\
\rule{0pt}{3ex}%
	$ m_{u} $ & /MeV & 0.602 & 0.351 &$\rightarrow$& 0.666 \\
	$ m_{c} $ & /MeV & 249.5 & 240.1 &$\rightarrow$& 257.5 \\
	$ m_{t} $ & /GeV & 93.37 & 89.84 &$\rightarrow$& 95.77 \\
	$ m_{d} $ & /MeV & 0.511 & 0.744 &$\rightarrow$& 0.929 \\
	$ m_{s} $ & /MeV & 15.80 & 15.66 &$\rightarrow$& 17.47 \\
	$ m_{b} $ & /GeV & 0.947 & 0.930 &$\rightarrow$& 0.953 \\[0.5ex]
\hline
	\multicolumn{2}{|c|}{\rule{0pt}{3ex}$ \delta\chi^2 $} & 16.0 &&&\multicolumn{1}{c}{} \\[0.5ex]
\cline{1-3}
\end{tabular}
\caption{Model predictions in the quark sector, for $ \tan \beta = 5 $. The quark contribution to the total $ \chi^2 $ is 16.0. Observables are at GUT scale.}
\label{tab:fitoutq}
\end{minipage}%
\qquad%
\begin{minipage}[b]{0.4\textwidth}
\captionsetup{width=0.8\textwidth}
\centering
\begin{tabular}{| c | c |}
\hline
\rule{0pt}{4.5ex}%
	Parameter	& Fitted value \\[2ex]
\hline\hline
\rule{0pt}{3ex}%
	$ y^u_\mathrm{atm} $ & 3.478 \exto{-5} \\
	$ y^u_\mathrm{sol} $ & 2.075 \exto{-4} \\
	$ y^u_\mathrm{dec} $ & 5.389 \exto{-1} \\
	$ y^u_\mathrm{sd} $ & 5.774 \exto{-3} \\
\rule{0pt}{3ex}%
	$ \eta^u $ & $ 1.629\pi $ \\
\rule{0pt}{3ex}%
	$ y^d_\mathrm{atm} $ & -3.199 \exto{-4} \\
	$ y^d_\mathrm{sol} $ & 2.117 \exto{-5} \\
	$ y^d_\mathrm{dec} $ & 2.792 \exto{-2} \\
	$ \eta $ & $ 2\pi/3 $ \\
\hline
\end{tabular}
\caption{Quark sector input parameter values (with $\eta$ fixed by the theory).}
\label{tab:fitinq}
\end{minipage}
\end{table}

\begin{table}
\centering
\footnotesize
\renewcommand{\arraystretch}{1.3}
\begin{minipage}[b]{0.5\textwidth}
\captionsetup{width=\textwidth}
\centering
\begin{tabular}{| c@{\hskip 4pt}l | c | r@{\hskip 3pt}c@{\hskip 3pt}l |}
\hline
\multicolumn{2}{|c|}{Observables} & \rule{0pt}{4.5ex}Model & \multicolumn{3}{c|}{\makecell{Data fit $ 1\sigma $ range \\ (from \cite{Antusch:2013jca})}} \\[2ex]
\hline\hline
\rule{0pt}{3ex}%
	$ \theta_{12}^l $ & /\degr & 33.13  & 32.83 &$\rightarrow$& 34.27 \\
	$ \theta_{13}^l $ & /\degr & 8.59 & 8.29 &$\rightarrow$& 8.68 \\
	$ \theta_{23}^l $ &/\degr & 40.81  & 40.63 &$\rightarrow$& 43.85 \\
\rule{0pt}{3ex}%
	$ \delta^l $ & /\degr & 280  & 192 &$\rightarrow$& 318 \\
\rule{0pt}{3ex}%
	$ m_{e} $ & /MeV & 0.342  & 0.340 & $\rightarrow$& 0.344 \\
	$ m_{\mu} $ & /MeV & 72.25  & 71.81 & $\rightarrow$& 72.68 \\
	$ m_{\tau} $ & /GeV & 1.229  & 1.223 & $\rightarrow$& 1.236 \\
\rule{0pt}{3ex}%
	$ \Delta m_{21}^2 $ & /eV$^2$ & 7.58 \exto{-5} & (7.33 & $\rightarrow$& 7.69) \exto{-5} \\
	$ \Delta m_{31}^2 $ & /eV$^2$ & 2.44 \exto{-3} & (2.41 & $\rightarrow$& 2.50) \exto{-3} \\
\rule{0pt}{3ex}%
	$ m_1 $ & /meV & 0.32 & & $ - $ & \\
\hline
	\multicolumn{2}{|c|}{\rule{0pt}{3ex}$ \delta\chi^2 $} & 1.3  &&& \\[0.5ex]
\hline
\end{tabular}
\caption{Model predictions in the lepton sector, for $ \tan \beta = 5 $. Observables are at GUT scale. }
\label{tab:fitoutl}
\end{minipage}%
\qquad%
\begin{minipage}[b]{0.4\textwidth}
\centering
\captionsetup{width=0.9\textwidth}
\begin{tabular}{| c | c |}
\hline
	Parameter	& Fitted value \\[1ex]
\hline\hline
\rule{0pt}{3ex}%
	$ y^e_\mathrm{dec} $ & 3.366 \exto{-2}  \\
	$ y^e_\mathrm{atm} $ & 2.217 \exto{-3}  \\
	$ y^e_\mathrm{sol} $ & -1.025 \exto{-5}  \\
\rule{0pt}{3ex}%
	$ \mu_\mathrm{dec} $ /meV & 2.052  \\
	$\mu_\mathrm{atm}$ /meV & 26.60  \\
	$ \mu_\mathrm{sol} $ /meV & 2.571  \\
\rule{0pt}{3ex}%
	$ \eta $ & $ 2\pi/3 $ \\
\hline
\end{tabular}
\caption{Lepton input parameter values (with $\eta$ fixed by the theory).}
\label{tab:fitinl}
\end{minipage}
\end{table}

\FloatBarrier
\section{Yukawa and Majorana parameters in terms of the fundamental model parameters\label{sec:yukawa}}
The Yukawa and Majorana mass matrices appearing in the seesaw Lagrangian in 
Eq.~\ref{eq:smm} are related to the real (due to $CP$ symmetry) fundamental model parameters by the following relations,
\begin{equation}
\medmuskip=0mu
\thinmuskip=0mu
\thickmuskip=2mu
\begin{split}
	(Y^\nu_{ij} )^*&= 
	\frac{\tilde\lambda^{(u)}_{\mathrm{atm}}}{ M_\chi^{3}}
	\braket{\overbar{\phi}^{i}_\mathrm{atm} \overbar{\phi}^{j}_\mathrm{atm} \xi}
	+ 
	\frac{\tilde\lambda^{(u)}_{\mathrm{sol}}}{ M_\chi^{4}}
	\braket{\overbar{\phi}^{i}_\mathrm{sol}\overbar{\phi}^{j}_\mathrm{sol} \xi^2}
	+
	\frac{\tilde\lambda^{(u)}_{\mathrm{dec}}}{ M_\chi^{2}}
	\braket{\overbar{\phi}^{i}_\mathrm{dec}\overbar{\phi}^{j}_\mathrm{dec}} ,
	\\
	(Y^e_{ij})^*&=
	\frac{\tilde\lambda^{(d)}_{\mathrm{atm}}}{ M_\chi^{4}}
	\braket{\overbar{\phi}^{i}_\mathrm{atm} \overbar{\phi}^{j}_\mathrm{atm} \xi^2}
	+ 
	\frac{\tilde\lambda^{(d)}_{\mathrm{sol}}}{M_\chi^{5}}
	\braket{\overbar{\phi}^{i}_\mathrm{sol} \overbar{\phi}^{j}_\mathrm{sol} \xi^3}
	+
	\frac{\tilde\lambda^{(d)}_{\mathrm{dec}}}{M_\chi^{3}}
	\braket{\overbar{\phi}^{i}_\mathrm{dec} \overbar{\phi}^{j}_\mathrm{dec}\xi},
	\\
	(M^N_{ij})^*&=
	\braket{H_{\overbar{16}}}^2
	\left[
	\frac{\tilde\lambda^{(M)}_{\mathrm{atm}}}{M_\chi^3 M_{\Omega_\mathrm{atm}}^4}
	\braket{\overbar{\phi}^{i}_\mathrm{atm} \overbar{\phi}^{j}_\mathrm{atm} \xi^4}
	+ 
	\frac{\tilde\lambda^{(M)}_{\mathrm{sol}}}{M_\chi^4 M_{\Omega_\mathrm{sol}}^4}
	\braket{\overbar{\phi}^{i}_\mathrm{sol} \overbar{\phi}^{j}_\mathrm{sol} \xi^5 }
	+
	\frac{\tilde\lambda^{(M)}_{\mathrm{dec}}}{M_\chi^2 M_{\Omega_\mathrm{dec}}^4}
	\braket{\overbar{\phi}^{i}_\mathrm{dec} \overbar{\phi}^{j}_\mathrm{dec} \xi^3}\right], 
\end{split}
\label{eq:Yflav}
\end{equation}
where the complex conjugation arises in going from the superpotential to the Lagrangian.
The structure of these Yukawa matrices is dictated by the complex VEVs of the flavons, which break $\Delta(27)$  and $CP$ symmetry in a specific way, giving the CSD3 alignment (see Eq.~\ref{eq:csd3alignments}).

The real parameters introduced in Eq.~\ref{eq:matrixstructure} can then be read off explicitly as follows:
{
\medmuskip=0mu
\thinmuskip=0mu
\thickmuskip=2mu
\begin{alignat}{6}
	y^\nu_\mathrm{atm} &= \frac{\tilde\lambda^{(u)}_\mathrm{atm} |v_\mathrm{atm}|^2 |v_\xi|}{M_\chi^3} , &\ \ 
	y^\nu_\mathrm{sol} &= \frac{\tilde\lambda^{(u)}_\mathrm{sol} |v_\mathrm{sol}|^2 |v_\xi|^2}{M_\chi^4} , &\ \ 
	y^\nu_\mathrm{dec} &= \frac{\tilde\lambda^{(u)}_\mathrm{dec} |v_\mathrm{dec}|^2          }{M_\chi^2} \nonumber \\
	y^e_\mathrm{atm} &= \frac{\tilde\lambda^{(d)}_\mathrm{atm} |v_\mathrm{atm}|^2 |v_\xi|^2}{M_\chi^4} , &
	y^e_\mathrm{sol} &= \frac{\tilde\lambda^{(d)}_\mathrm{sol} |v_\mathrm{sol}|^2 |v_\xi|^3}{M_\chi^5} , &
	y^e_\mathrm{dec} &= \frac{\tilde\lambda^{(d)}_\mathrm{dec} |v_\mathrm{dec}|^2 |v_\xi|  }{M_\chi^3} \nonumber \\
	M_\mathrm{atm} &= \frac{\tilde\lambda^{(M)}_\mathrm{atm} |v_\mathrm{atm}|^2 |v_\xi|^4 |v_{H_{\overbar{16}}}|^2}{M_\chi^3 M_{\Omega_\mathrm{atm}}^4} , & 
	M_\mathrm{sol} &= \frac{\tilde\lambda^{(M)}_\mathrm{sol} |v_\mathrm{sol}|^2 |v_\xi|^5 |v_{H_{\overbar{16}}}|^2}{M_\chi^4 M_{\Omega_\mathrm{atm}}^4} , &
	M_\mathrm{dec} &= \frac{\tilde\lambda^{(M)}_\mathrm{dec} |v_\mathrm{dec}|^2 |v_\xi|^3 |v_{H_{\overbar{16}}}|^2}{M_\chi^2 M_{\Omega_\mathrm{atm}}^4} , \nonumber \\
	\eta &= -\arg\left[\frac{v_\mathrm{sol}^2}{v_\mathrm{atm}^2} v_\xi \right] , & 
	\eta^\prime &= -\arg\left[\frac{v_\mathrm{dec}^2}{v_\mathrm{atm}^2} \frac{1}{v_\xi} \right] .&&
\label{eq:ydefinitions}
\end{alignat}
}

\section{The seesaw mechanism with rank-one matrices \label{sec:seesaw}}
Defining three rank-one matrices in terms of a set of column vectors ${\phi}_a $, ${\phi}_b $, ${\phi}_c$,
\begin{equation}
	A=\phi_a\phi_a^T, \ \ B=\phi_b\phi_b^T, \ \ C=\phi_c\phi_c^T,
\end{equation}
the Dirac mass matrix $m_D$ and heavy RH Majorana matrix $M_R$ may be written as 
\begin{equation}
\begin{split}
	m_D &= m_a A + m_b B + m_c C,  \\
	M_R &= M_a A + M_b B + M_c C.
\end{split}
\label{eq:rank3}
\end{equation}
Now let us consider a new set of column vectors $\tilde{\phi}_a $, $\tilde{\phi}_b $, $\tilde{\phi}_c$, which are orthogonal to the original ones, and satisfy the conditions
\begin{equation}
	\tilde{\phi}_i ^T  \phi_j = \delta_{ij} , \ \ i,j=a,b,c.
\label{eq:orthog}
\end{equation}
For example, for the column vectors used in this paper,
\begin{equation}
	\phi_a=(0,1,1)^T, \quad \phi_b=(1,3,1)^T, \quad \phi_c=(0,0,1)^T.
\end{equation}
the corresponding column vectors which satisfy Eq.~\ref{eq:orthog} are,
\begin{equation}
	\tilde{\phi_a}=(-3,1,0)^T, \quad \tilde{\phi_b}=(1,0,0)^T, \quad \tilde{\phi_c}=(2,-1,1)^T.
\end{equation}
Given these new vectors, we can define some new rank-one matrices,
\begin{equation}
	 \tilde{A} = \tilde{\phi}_a \tilde{\phi}_a^T , \quad  
	 \tilde{B} = \tilde{\phi}_b \tilde{\phi}_b^T, \quad 
	 \tilde{C} = \tilde{\phi}_c \tilde{\phi}_c^T. 
\label{eq:At}
\end{equation}
Then the inverse of the heavy RH Majorana matrix is uniquely given as
\begin{equation}
	M_R^{-1}= \frac{1}{M_a}\tilde{A}+\frac{1}{M_b}\tilde{B}+\frac{1}{M_c}\tilde{C}.
\label{MRI}
\end{equation}
It can easily be verified explicitly that this result satisfies $M_RM_R^{-1}=I$ using Eq.~\ref{eq:orthog}, which implies that the cross-terms vanish, e.g. $A \tilde{B} =0 $.
It is also worth noting that the orthogonality condition in Eq.~\ref{eq:orthog} is sufficient for immediately computing the unique solution for inverse. As any rank-three (inverse) matrix can be written as the sum of three rank-one matrices (such as in Eq.~\ref{MRI}), the orthogonality condition arises when we require $ M_R^{-1} M_R $ to be independent of the scaling factors $ 1/M_i $.

Using Eq.~\ref{MRI} we can explicitly evaluate the seesaw formula, $m^{\nu}=m_DM_R^{-1}m_D^T $,
\begin{equation}
	m^{\nu} = \frac{m_a^2}{M_a}A + \frac{m_b^2}{M_b}B + \frac{m_c^2}{M_c}C 
\label{eq:mnuseesaw2}
\end{equation}
where, as before, we have used Eq.~\ref{eq:orthog}, which implies cross-terms vanish and $A\tilde{A} A = A $, and so on, from which we see that 
\begin{equation}
	m^{\nu}=\mu_aA + \mu_bB + \mu_cC
\end{equation}
where $\mu_a= m_a^2/M_a$, $\mu_b= m_b^2/M_b$, $\mu_c= m_c^2/M_c$.
Clearly this result is valid for any choice of linearly independent column vectors ${\phi}_a $, ${\phi}_b $, ${\phi}_c$.

\end{document}